\documentclass[prl,10pt,superscriptaddress,twocolumn]{revtex4}

\usepackage{graphicx}
\usepackage{amsmath}
\usepackage{color}
\usepackage{units}
\usepackage[latin1]{inputenc}
\usepackage{verbatim}
\usepackage{float}
\usepackage[english]{babel}
\usepackage{booktabs}	
\usepackage{epstopdf}
\usepackage{pgf}
\usepackage{pgfplots}
\usepackage{placeins}

\newcommand{\tsub}[2]{\ensuremath{\text{#1}_\text{#2}}}

\begin{document}


\title{Sensitivity improvement of a laser interferometer limited by inelastic back-scattering, employing dual readout}

\author{Melanie\,Meinders}
\affiliation{Institut f\"ur Laserphysik und Zentrum f\"ur Optische Quantentechnologien der Universit\"at Hamburg,\\%
Luruper Chaussee 149, 22761 Hamburg, Germany}
\affiliation{Institut f\"ur Gravitationsphysik der Leibniz Universit\"at Hannover and\\ %
Max-Planck-Institut f\"ur Gravitationsphysik (Albert-Einstein-Institut), Callinstra{\ss}e 38, 30167 Hannover, Germany}%
\author{Roman Schnabel}
\email[corresponding author:\\]{roman.schnabel@physnet.uni-hamburg.de}
\affiliation{Institut f\"ur Laserphysik und Zentrum f\"ur Optische Quantentechnologien der Universit\"at Hamburg,\\%
Luruper Chaussee 149, 22761 Hamburg, Germany}
\affiliation{Institut f\"ur Gravitationsphysik der Leibniz Universit\"at Hannover and\\ %
Max-Planck-Institut f\"ur Gravitationsphysik (Albert-Einstein-Institut), Callinstra{\ss}e 38, 30167 Hannover, Germany}%

\begin{abstract} 
Inelastic back-scattering of stray light is a long-standing and fundamental problem in high-sensitivity interferometric measurements and a potential limitation for advanced gravitational-wave detectors. The emerging parasitic interferences cannot be distinguished from a scientific signal via conventional single readout. In this work, we propose the subtraction of inelastic back-scatter signals by employing dual homodyne detection on the output light, and demonstrate it for a table-top Michelson interferometer. The additional readout contains solely parasitic signals and is used to model the scatter source. Subtraction of the scatter signal reduces the noise spectral density and thus improves the measurement sensitivity. Our scheme is qualitatively different from the previously demonstrated vetoing of scatter signals and opens a new path for improving the sensitivity of future gravitational-wave detectors  and other back-scatter limited devices. 

\end{abstract}
\maketitle

\begin{figure}[p b]
 \center 
  \includegraphics[width=\columnwidth]{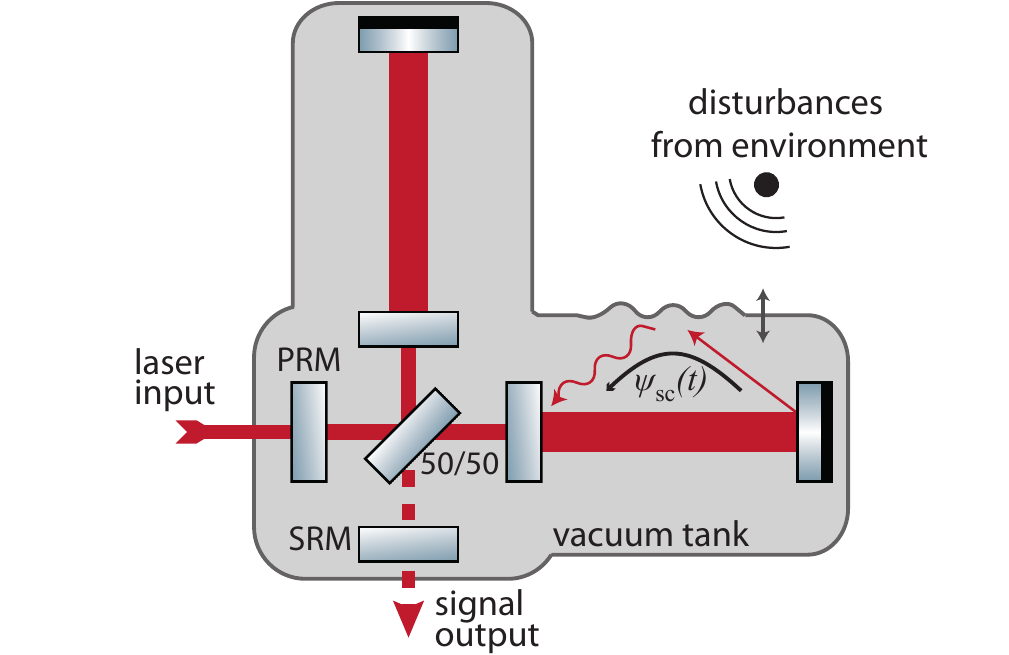}
  \caption{\textbf{Scenario for the occurrence of inelastic back-scattering in gravitational-wave detectors. }The picture shows the advanced detector topology with arm cavities and power- and signal-recycling mirrors (PRM, SRM), placed in vacuum. Light is scattered out of the interferometer mode, e.g.~due to micro-roughness of mirror surfaces, and accumulates a time-dependent phase shift $\psi_\text{sc}(t)$ due to changes of the optical path length. Back-scattering into the interferometer mode produces a disturbance signal at the output.}
  \label{Fig1}
\end{figure}

Parasitic signals, arising from inelastic back-scattering of stray light, are a recurrent issue in high-sensitivity laser interferometers like gravitational-wave (GW) detectors. These detectors use intense laser light to measure differential arm length changes in a Michelson interferometer topology, reaching strain sensitivities in the order of $10^{-22}\unit{/\sqrt{Hz}}$ \cite{LIGO2009}. 
The scattered light problem was discovered right in the beginning in the first prototypes \cite{Bill1979} and has been observed in all first-generation detectors \cite{Hild2006}. An overview on this topic can be found in \cite{AdvDetBook}. A sketch of a possible scatter scenario is depicted in Fig.~\ref{Fig1}. Shown is the advanced GW-detector topology that includes power recycling, (tuned) signal recycling and arm cavities, placed in a vacuum chamber. Imperfect anti-reflection coatings of transmissive optics and micro roughness of mirror surfaces lead to stray light. Small amounts are back-reflected from the surroundings, in this example, from the walls of the vacuum tank whose motion is excited by acoustic or seismic disturbances from the environment. Changes in the optical path length cause phase modulations of the stray light, which are associated with changes of the optical frequency. Due to this inelastic character of the scattering, the recombination with the interferometer mode then produces a disturbance signal in the output light. For motions at audio frequencies, these scatter signals show up directly in the detector's most sensitive band. Even worse are sources that move at lower frequencies but with large motional amplitudes of several wavelengths, e.g.~due to micro-seismic. In that case, frequency up-conversion leads to broadband scatter shoulders that can completely cover the most interesting gravitational-wave signal band. Observations of this kind were described for example for VIRGO's second science run \cite{Fiori2010}. 
For mitigation, the two main approaches are to reduce the amount of stray light as far as possible and to improve the isolation of potential scatter sources from the environment to reduce their motion. Clearly, technical feasibility limits both methods. In addition, future increase of the light power for shot-noise reduction \cite{formula} will make the detectors even more vulnerable to inelastic back-scattering.  {Also, the extension towards lower frequencies ($<10\,\unit{Hz}$) in third generation detectors will require significantly improved mitigation schemes against back-scattered light \cite{Ottaway2012}.} 

In this work we propose and demonstrate a new approach, complementing the existing mitigation techniques. We collect information about the scatter signal independently from the scientific  phase signal, using the orthogonal (amplitude) quadrature. Up to now, such additional information has been used solely for vetoing of disturbance signals \cite{Ball2009,SBM2013}, i.e.~excluding corrupted measurement data from scientific analysis. In our approach, we use the data from the amplitude measurement $x(t)$ to calculate a time-dependent phase-space model of the scattered light $\psi_\text{sc}(t)$ and subtract its projection from the phase measurement data $p(t)$. We thereby restore sensitivity in the affected frequency range.\\ 
%


Stray light travels an extra path with respect to the interferometer mode and generally leads to phase as well as amplitude modulations of the interferometer output. 
A displacement of the back-scatter source changes the optical path length $s(t)$ and results in a time-dependent phase shift $\psi_\text{sc}(t)=\tfrac{2\pi}{\lambda}s(t)$ for the scattered light. Here, $\lambda$ denotes the laser wavelength. The projections of the scatter signal into the amplitude and phase quadratures $x(t)$ and $p(t)$ are given by
\begin{equation}\label{eqproj}
 {x_\text{sc}(t)=A\,\sin\psi_\text{sc}(t) \quad\text{and} \quad p_\text{sc}(t)=A\,\cos\psi_\text{sc}(t)\,,}
\end{equation}%
where the back-scatter signal amplitude $A$ for instance depends on the transfer function of the interferometer and on the intensity of the scattered light. The interferometer's amplitude quadrature ($x(t)$) readout does not contain any GW-signals. If the laser input light is well-stabilized or the interferometer has equal arm lengths and is operated at a dark fringe, this readout solely contains back-scatter signals and shot-noise. 
It is thus an unbiased monitor for the scatter signal from which the back-scatter signal amplitude $A$ and the time dependent phase shift $\psi_\text{sc}(t)$ can be extracted by fitting an analytical model of the back-scatter source to the $x(t)$ data. The projection into the phase quadrature ($p(t)$) can be calculated according to the right side of Eq.~(\ref{eqproj}).

In this work, we consider the case of a scatter shoulder that is produced by back-scatter sources with large motional amplitudes, like already mentioned above and described in \cite{Bill1979, Hild2006, AdvDetBook, Fiori2010}. To keep the analytical model simple, let us consider a single source that is moving sinusoidally at a constant average distance to the interferometer. This results in a modulated optical path length
\begin{equation}\label{eqdispl}
s(t)=s_0+\,m\,\sin(2\pi\,f_\text{m}\,t+\varphi_\text{m})\,
\end{equation}%
with a constant average path length $s_0$, modulation depth $m$, frequency $f_\text{m}$ and phase $\varphi_\text{m}$. The signal amplitude $A$ of Eq.~(\ref{eqproj}) is considered to be frequency independent and also constant over the time measured. The Doppler shift of the back-scattered light is proportional to the change of the optical path length $f_\text{ds}(t)=\tfrac{\dot{s}(t)}{\lambda}$, which results in:%
\begin{equation} \label{eqfds}
f_{\text{ds}}(t)=\frac{2\pi}{\lambda}\,m\,f_\text{m}\,\cos(2\pi\,f_\text{m}\, t+\varphi_\text{m}).
\end{equation}%
The maximum frequency component as observed in a single-sided spectrum is then given by $f_\text{ds}^\text{max}=|\tfrac{2\pi}{\lambda}\,m\,f_\text{m}|$ and for a modulation depth $m>\lambda/2\pi$, this represents frequency up-conversion.

Although the model described here is quite simple, it already reproduces the basic structure of typical disturbance signals as observed in GW-detectors. It does generate a broadband `shoulder', which even includes an overlaid bump structure, as observed in the LIGO data presented in Fig.\,3.1 of reference \cite{Hild2006}. This might indicate that indeed slowly varying mechanical vibrations give rise to a significant share of the back-scattered light disturbances observed in GW-detectors. Our technique is not restricted to a simple sinusoidal model, in fact, our experimental demonstration actually required a model that included higher harmonics in the motion of the back-scatter source.

\begin{figure}[b]
 \center
  \includegraphics[width=\columnwidth]{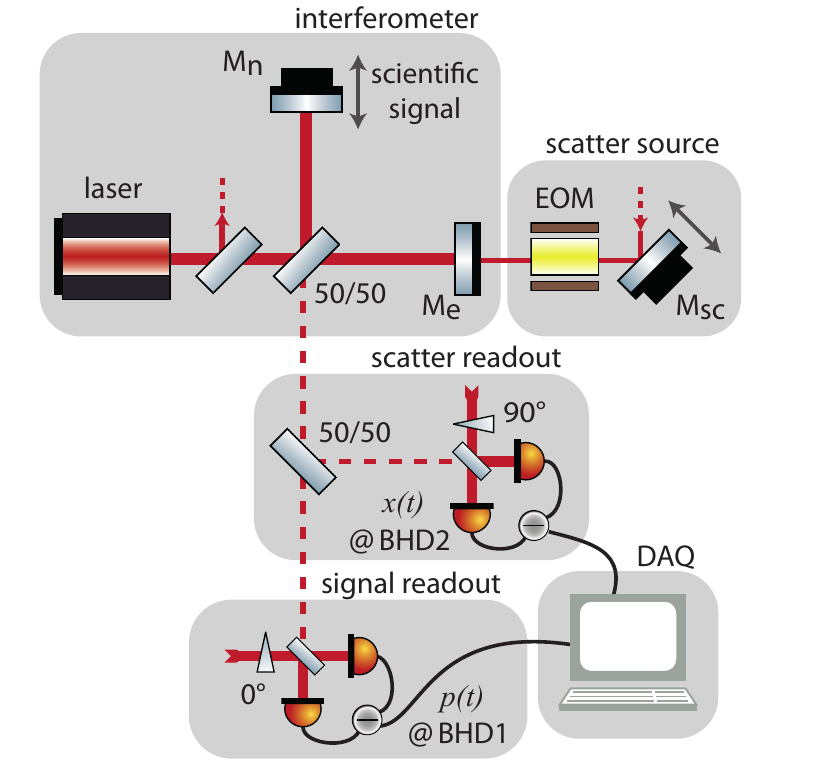}
  \caption{\textbf{Schematic of the experimental setup. }In our table-top Michelson interferometer two signals were being generated: A scientific signal (pure phase modulation) in the north arm and a parasitic signal (broadband scatter shoulder, appearing in phase \emph{and} amplitude) in the east arm with the help of an electro-optic modulator (EOM) and a piezo actuated mirror \tsub{M}{sc}. The output signal was split at a 50/50 beam splitter. Two balanced homodyne detectors (BHD1\&2) measured the orthogonal quadratures  {($p(t)$,\,$x(t)$)}.  {`$0^\circ$'\,refers to the local oscillator phase that is optimal for the scientific signal. DAQ: data acquisition system.}
} 
\label{Fig2}
\end{figure}
\begin{figure}[h!bp]
 \center
\includegraphics[width=\columnwidth]{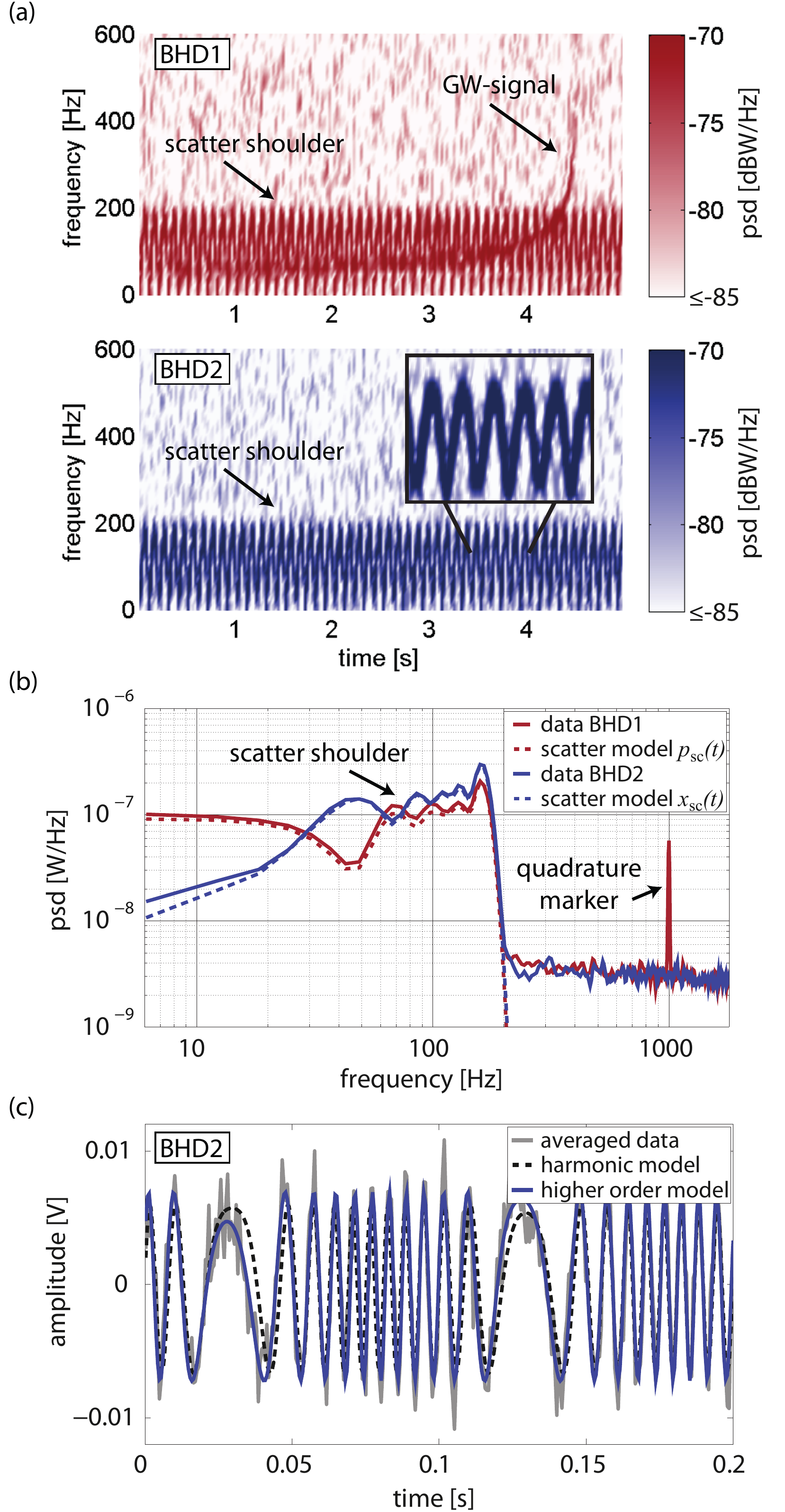}
\caption{ \textbf{Dual readout data (demodulated at 5.2\,MHz).} (a) Spectrogram of the measurement data from BHD1\&2 (red\,\&\,blue) showing the time resolved scatter shoulder (arches) concealing the simulated gravitational wave signal (inspiral) that was injected from an audio file \cite{gwsound}. (b) Averaged spectrum showing the scatter shoulder limiting both quadrature measurements over a  {bandwidth of about 0.2\,kHz}. Apart from this, the measurements were limited by optical shot noise. (c) Original time data of BHD2 (averaged) with two fitted scatter models. A higher order model provides the better fit because it takes the piezo nonlinearity into account.}
 \label{Fig3}
\end{figure}

%
In the experiment, we prepared a table-top Michelson interferometer that was limited by inelastic back-scattering over a broad frequency range, as described below. The setup is depicted in Fig.~\ref{Fig2}. The interferometer had an arm length of about 7\,cm and an input power of 10\,mW at a wavelength of $\lambda=1064\,\unit{nm}$. It was stabilized to a dark fringe using the Schnupp-modulation technique \cite{Schnupp}. To artificially produce a scatter shoulder, according to Eq.~(\ref{eqproj})-(\ref{eqfds}), we picked off a low power light beam in front of the interferometer and injected it through the back of end mirror \tsub{M}{e} in the east arm. We modulated its phase with a piezo actuated mirror \tsub{M}{sc}, driven sinusoidally at a frequency of 5\,Hz and with a modulation depth of a few $\lambda$ {, which resulted in a  {maximum frequency shift of $f_\text{ds}^\text{max}\approx 0.2\,\unit{kHz}$}.} The center of motion of the mirror \tsub{M}{sc} was not stabilized but turned out to be sufficiently constant over the measurement time. An additional phase modulation at 5.2\,MHz was imprinted by an electro-optic modulator (EOM) to shift the scatter signal into the MHz range. At these frequencies our measurements were generally limited by optical shot-noise and we could avoid additional disturbances, e.g.~due to acoustics. Note, that the light was picked off outside the interferometer because this way it was easier to produce sufficiently large scatter signals in our table-top experiment. We also introduced a scientific signal due to a real differential arm length change, to show that these kind of signals are not affected by the subtraction of the scatter model. The scientific signal we generated by modulating the piezo mounted end mirror \tsub{M}{n} in the north arm of the interferometer with a sound file \cite{gwsound}. The file contained about 4.5 seconds of a simulated inspiral of two neutron stars with equal masses. The sound file was fed into an Agilent 33500B series waveform generator as an external modulation and shifted by 5.2\,MHz before it was put on the piezo. To measure the  {phase and amplitude} quadratures simultaneously, the output signal was split up at a 50/50 beam splitter. This resulted in 50\% loss for the scientific signal.  {Generally, other splitting ratios can be used, compromising between loss for the scientific signal and signal-to-shot-noise-ratio in the scatter monitor.}  
The outputs of the beam splitter were read out with two balanced homodyne detectors (BHD1\&2) each using a local oscillator power of about 8\,mW. The detectors were stabilized $90\,^\circ$\,out of phase and the phase space orientation with respect to the interferometer signal was adjusted by minimizing a marker peak at 5.2\,MHz+1\,kHz in the live spectrum of BHD2. The marker was also generated with the piezo actuated end mirror \tsub{M}{n}. The detected signals were mixed down at 5.2\,MHz to recover the audio-band signals and sent through an anti-aliasing filter. They were then acquired with a PCI-6259 card from National Instruments and processed in LabView. The post-processing was done in Matlab. The measurement results, employing the dual readout for the scatter limited interferometer, are depicted in Fig.~\ref{Fig3}.


Fig.~\ref{Fig3}\,(a) shows the measurements from both detectors in spectrograms to cope with the time dependence of the injected signals. The scatter signal shows up as `arches' in these plots, which are more clearly visible in the zoomed-in cutout of the data from BHD2 (blue). The arches follow from the absolute value of the time-dependent frequency shift of Eq.~(\ref{eqfds}). The injected GW-signal is almost completely covered by the scatter signal, only its `tail' towards high frequencies is clearly visible in the data of BHD1 (red). 
In the averaged spectrum of Fig.~\ref{Fig3}\,(b) the scatter signal appears as the name-giving, broadband shoulder in both phase (BHD1, solid red) and amplitude (BHD2, solid blue) quadrature. Its bump structure results from the projection into the quadratures because the different frequency components are being generated at different total distances from the interferometer. The averaged power spectral density (psd) was computed with Matlab's 'pwelch' function, using a Hanning window spanning half the oscillation period of the scatter source ($\Delta t=1/(2f_\text{m})$)  and an overlap of 50\%. These settings were chosen to get a sufficient frequency resolution without reducing the shoulder to the harmonics of $f_\text{m}$ by temporal averaging. The scatter shoulder is clearly the dominant noise source for frequencies below 0.2\,kHz, whereby this frequency refers to a demodulated frequency from the MHz regime.  {Above 0.2\,kHz} the measurements were limited by optical shot-noise. The peak at 1\,kHz in the spectrum of BHD1 shows the earlier mentioned marker, used to determine the quadratures. The injected GW-signal can not be identified in this plot, only a slight rise of the noise floor at the edge of the scatter shoulder of BHD1 is visible. 

In the attempt to model the scatter signal measured at BHD2 with Eq.~(\ref{eqproj})\,\&\,(\ref{eqdispl}) it turned out that the harmonic model for the source displacement in Eq.~(\ref{eqdispl}) was not sufficient. An averaged section of the recorded time data from BHD2 is depicted in Fig.~\ref{Fig3}\,(c) (solid gray), together with the obtained fit (dashed black). Especially around the turning points of the piezo at about 0.03\,s and 0.13\,s, there is a quite strong deviation between the harmonic model and the measurement data. To account for the nonlinear response of the piezo to the incoming sine wave, we included higher harmonics of the scatterer's oscillation frequency in the model for the displacement
\begin{equation}\label{eqNLdispl}
s(t)=\sum_{n=0}^{5}{m_n\,\sin(2\pi\,f_\text{m}\,t+\varphi_{\text{m},n})^n} ,
\end{equation}
which leads to the computed fit given by the solid blue line in Fig.~\ref{Fig3}\,(c). For the fitting, no prior knowledge about the parameters of the scatter source was assumed. For the projection of the scatter model to the  {phase quadrature}, we allowed the signal amplitudes at the two detectors to differ by a constant factor, to compensate for example for an unbalanced splitting. Also, an additional constant phase was added in the cosine of Eq.~(\ref{eqproj}) to compensate for a non-perfect quadrature orientation of the detectors.  The two new parameters were fitted using the data of BHD1, while all parameters obtained in the fit of the  {amplitude} quadrature were kept fixed. The final models for both quadratures are given by the dashed lines of the respective colors in Fig.~\ref {Fig3}\,(b).

\begin{figure}[tp]
 \center
\includegraphics[width=\columnwidth]{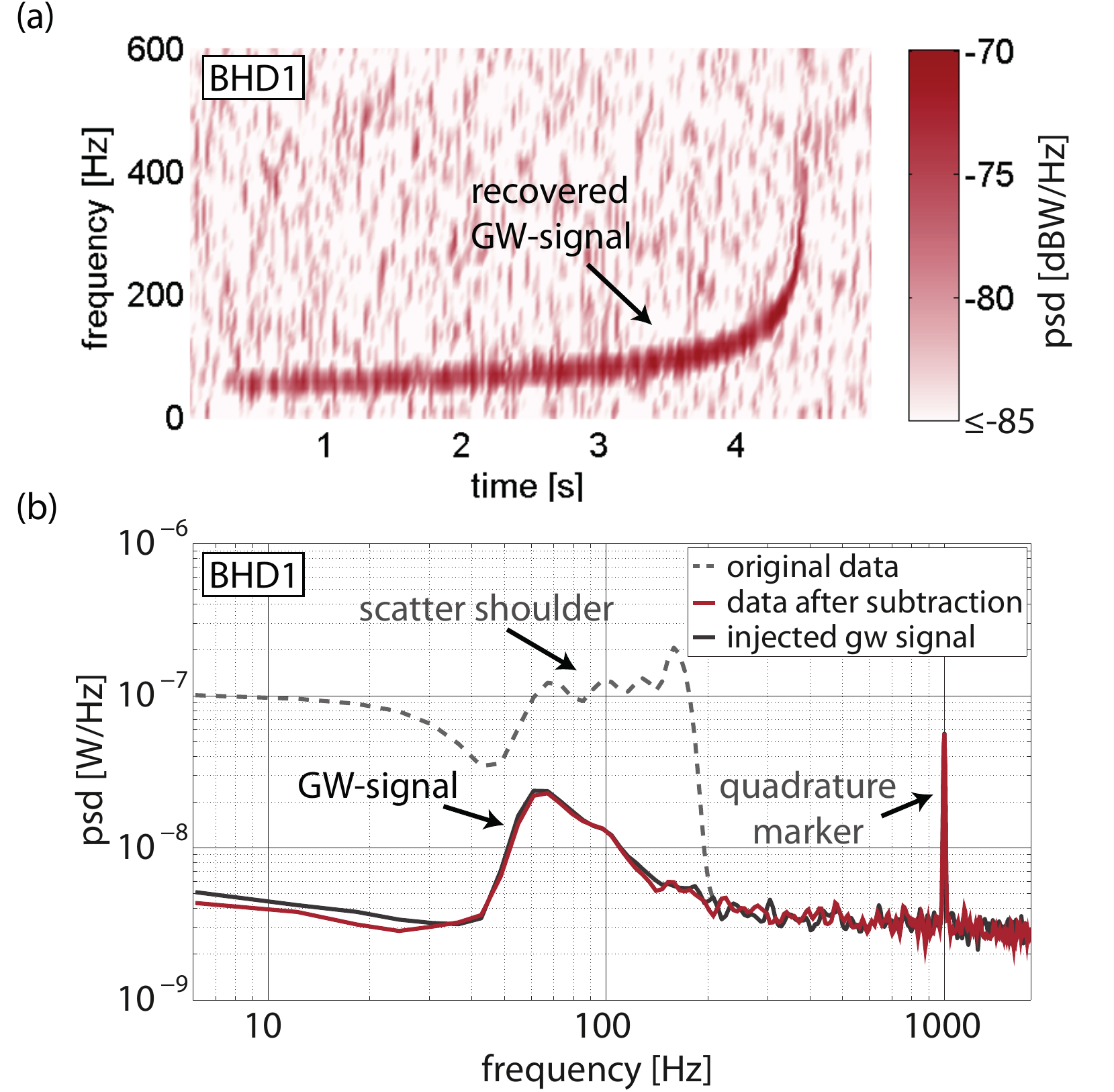}
\caption{ \textbf{Results after subtraction of the scatter model. }(a) Spectrogram of the measurement data from BHD1 after subtraction of the scatter model. The simulated gravitational wave signal \cite{gwsound} was clearly recovered. (b) Averaged spectrum showing the recovered GW-signal after subtraction of the scatter model (solid red) in comparison with the original data (dashed gray) and a reference measurement where only the GW-signal was injected and no scattering (solid dark gray). }
 \label{Fig4}
\end{figure}

The results for the subtraction of the modeled inelastic back-scattering (in time domain) from the  {phase} measurement at BHD1 are presented in Fig.~\ref{Fig4}.  {It is clearly visible in the spectrogram of Fig.~\ref{Fig4} (a) that the injected GW-signal could be fully recovered. The arches are not visible anymore.} In Fig.~\ref{Fig4} (b) the subtracted data (solid red) is compared to a reference measurement (solid dark gray) recorded while the scattering was blocked and only the simulated GW-signal was being injected. For comparison, the dashed gray line in Fig.~\ref{Fig4} (b) shows the original measurement data. In the  {scatter limited frequency range below 0.2\,kHz}, a sensitivity improvement of more than one order of magnitude was achieved, with a final sensitivity limited by optical shot-noise. 

To conclude, we present a proof-of-principle experiment, in which we show for the first time, to the best of our knowledge, that the noise spectral density of a measurement device can be reduced by using an additional readout that detects the observable that is orthogonal to the one that carries the scientific signal. Our result is thus of high relevance for metrology in general. We explicitly show that a broadband disturbance spectrum due to parasitic inelastic back-scattering can be removed using the additional data. 
The balanced splitting of our dual readout increases the signal-normalized shot noise by a factor of 2 in power. This disadvantage can be mitigated by reducing the amount of power tapped for the scatter readout, which in turn should result in a reduced quality of the back-scatter subtraction. Compromising the two effects requires further investigations. 
The increased shot noise is not an issue at all, however, if the overall noise is dominated -- apart from back-scatter -- by e.g.~quantum back-action or thermal noise. Since a shot-noise limited sensitivity is usually only achieved above some corner frequency, a carrier-tuned optical cavity could be used as a filter for optimizing our dual readout. Sideband frequencies above the filter linewidth are reflected from the filter and should be analyzed by a conventional single readout, while the transmitted lower frequencies should be analyzed by the dual readout.

In principle our scheme can directly be used in future GW detectors. The additional readout requires an optical local oscillator for balanced homodyne detection, which has not been used in GW detectors so far, but was already suggested for future sensitivity improvements \cite{KLMV2001}. Presumably, all optics that guide the local oscillator and the interferometer output field need to be suspended to avoid differential phase fluctuations. 

Our result is not restricted to GW detectors, but can be seen as a fundamental approach to remove measurement noise that also reveals itself in the observable that does not commute with the actual measurement quantity.\\

\FloatBarrier
\begin{acknowledgments}
This work was supported by the Deutsche Forschungsgemeinschaft (Sonderforschungsbereich Transregio 7, project C8) and by the International Max Planck Research School for Gravitational Wave Astronomy (IMPRS-GW).
\end{acknowledgments}




\begin{thebibliography}{99}
\newcommand{\enquote}[1]{``#1''}
\bibitem{LIGO2009} B. P. Abbott \textit{et al.}, LIGO: the laser interferometer gravitational-wave observatory.
\textit{Rep. Prog. Phys.} \textbf{72}, 076901 (2009).
\bibitem{Bill1979} H. Billing, K. Maischberger, A. R\"udiger, R. Schilling, L. Schnupp, L. Winkler, An argon laser interferometer for the detection of gravitational radiation. \textit{J. Phys. E: Sci. Instrum.} \textbf{12}, 1043 (1979)
\bibitem{Hild2006} S. Hild, Beyond the First Generation: Extending the Science Range of the Gravitational Wave Detector GEO600. PhD Thesis, 31-32 (2006)
\bibitem{AdvDetBook} M. Bassan (ed.), Advanced Interferometers and the Search for Gravitational Waves. Astrophysics and Space Science Library 404, DOI: 10.1007/978-3-319-03792-9\_1, Springer International Publishing Switzerland, 275-290 (2014)
\bibitem{Fiori2010} I. Fiori \textit{et al.}, Noise from scattered light in Virgo's second science run data. \textit{Class. Quantum Gravity} \textbf{27}, 194011 (2010)
\bibitem{formula} The (back-action free) quantum-noise linear spectral density of a laser interferometer in $\rm m/\sqrt{\rm Hz}$ is proportional to $\text{e}^{-r}\sqrt{\hbar c \lambda/(2\pi P)}$, where $P$ is the light power inside the interferometer, $r$ the squeezing parameter and $\lambda$ the light's wavelength.
\bibitem{Ottaway2012} D. J. Ottaway, P. Fritschel and S. J. Waldman, Impact of upconverted scattered light on advanced interferometric gravitational wave detectors, \textit{Opt. Express} \textbf{20}, Issue 8, 8329-8336 (2012)
\bibitem{Ball2009} T. Ballinger (for the LIGO Scientific Collaboration and the Virgo Collaboration), A powerful veto for gravitational wave searches using data from Virgo's first scientific run. \textit{Class. Quantum Grav.} \textbf{26}, 204003 (2009)
\bibitem{SBM2013}  S. Steinlechner, J. Bauchrowitz,  M. Meinders, H. M\"uller-Ebhardt, K. Danzmann and R. Schnabel, Quantum-dense metrology. \textit{Nat. Phot.} \textbf{7},  626-630 (2013)
\bibitem{Schnupp} L. Schnupp, Presentation at the European Collaboration Meeting on Interferometric Detection of Gravitational Waves (Sorrent, 1988)
\bibitem{gwsound} Simulated sound of binary neutron stars from the research group of  Professor Scott A. Hughes at MIT, http://gmunu.mit.edu/sounds/comparable\_sounds/\\comparable\_sounds.html
\bibitem{KLMV2001} H. J. Kimble \textit{et al.}, Conversion of conventional gravitational-wave interferometers into quantum nondemolition interferometers by modifying their input and/or output optics, \textit{Phys. Rev. D} \textbf{65}, 022002 (2001). 


\end{thebibliography}
\end{document}